\documentstyle[amssymb,aps]{revtex}

\begin{document}
\title{The Many-Body Diffusion Algorithm, Harmonic Fermions}
\author{F. Luczak, F.~Brosens 
\thanks{Senior Research Associate of the FWO-Vlaanderen.}%
\thinspace and J.T. Devreese 
\thanks{Also at the Universiteit Antwerpen (RUCA) and Technische Universiteit Eindhoven, The Netherlands.}%
}
\address{Departement Natuurkunde, Universiteit Antwerpen (UIA),\\
Universiteitsplein 1, B-2610 Antwerpen}
\author{L. F. Lemmens}
\address{Departement Natuurkunde, Universiteit Antwerpen (RUCA),\\
Groenenborgerlaan 171, B-2020 Antwerpen}
\date{}
\maketitle

\begin{abstract}
A numerical implementation scheme is presented for the recently developed
many-body diffusion approach for identical particles, in the case of
harmonic potentials. The procedure is free of the sign problem, by the
introduction of the appropriate absorption or reflection conditions for the
walkers at the boundary of a state space. These conditions are imposed by
the permutation symmetry. The outflow of the walkers at the boundary of the
state space contributes substantially to the energy. Furthermore, the
implementation of crossing/recrossing effects at absorbing boundaries proves
indispensable to sample the antisymmetric states by discrete time steps.
\end{abstract}

\section{Introduction}

In this paper the theoretical method put forward in a series of publications
by three of the present authors \cite{BDL94,BDL95,BDL96} around the
criticism \cite{REPLY} on an approach initiated by Korzeniowski et al \cite
{KORZE} is applied to an analytically exact soluble model. A
sign-problem-free estimation of an excited antisymmetric eigenstate of the
model is presented.

Monte Carlo simulations of an excited state of quantum systems are only
possible if symmetry considerations can be invoked to guarantee that the
excited state is orthogonal to all lower lying states of the same system.
Usually this means that one simulates the state with the lowest energy that
transforms according to a particular irreducible representation of the
symmetry group of the system. Usually this also means that the state has a
nodal structure in the configuration space. This is e.g. the case if one
wants to simulate a system containing a fixed number of fermions. The
(unknown) nodal structure of this system is the origin of the so-called sign
problem. The fermion state to be simulated is an excited state of a quantum
system of distinguishable particles (the walkers in the simulation are
registered) that transforms according to the antisymmetric irreducible
representation of the permutation group $S_N,$ where $N$ is the number of
indistinguishable particles. Up till now, there is no generally accepted
way, that solves the sign problem. Several approaches avoid the problem
approximately by fixing the nodal surfaces to be those of a known, usually
exactly solvable problem, or by allowing that parts of the configuration
space with different sign are sampled in combination with variance reducing
methods \cite{SCH84,FIXNO,PROJE}.

Oscillators belong to the class of simple models that can be tested easily 
\cite{TAK84,NEW92,HAL92} and gain importance in view of the models for
trapped systems \cite{BUT97}. The ground-state energy and the static
response functions of particles in a harmonic potential, can be calculated
with their statistics taken into account, even in the presence of harmonic
two-body interactions. It is clear that analytic path-integral methods \cite
{BDLPRE97a,BDLPRE97b,BDLPRA97} allow the study of these models in a more
direct way. But it is not without value in view of the sign problem if the
lowest energy state of this harmonic system with a specified symmetry can be
simulated with a sign-problem-free procedure. Such a procedure will be
applied in the present paper. We want to stress that our only inference is
that we simulate the model (fermions interacting harmonically and
isotropically) sign-problem-free. The adaptation of the technique for a
realistic model is beyond the scope of the present communication. The
limitation of the proposed method originates from the fact that in the
theoretical framework \cite{BDL94,BDL95,BDL96} leading to this algorithm, we
rely on a typical symmetry property of harmonic interactions, which is
absent for more general potentials \cite{BDLer97}. Nevertheless, the basic
idea of using the permutation symmetry to partition the configuration space
in domains with boundary conditions offers the potentiality to lead to a
general procedure for more realistic systems.

From our theoretical contributions to this problem \cite{BDL94,BDL95,BDL96},
it should be clear that, in an algorithmic way, we have to study a new
problem, formerly not encountered in the area of quantum Monte Carlo
simulations and relating the sign problem to absorption and reflection of
walkers at a boundary. The novelty is that the boundary for a walker depends
on the positions of some other walkers.

The introduction of a domain also has consequences for the estimation of the
ground-state energy. With ground-state energy we mean here the lowest energy
of all the states that transform according to a specified irreducible
representation of a symmetry group. The absorption of walkers at the
boundary of the domain creates a flux of walkers out of the phase space
volume that gives an important contribution to the estimate for the energy
as will be shown below.

This paper is organized as follows. In section II we will discuss what
absorption or reflection means for the distribution of particles in a domain 
$D_N^{\,3}$ and how this distribution can be simulated by Bernoulli walks or
by diffusion Monte-Carlo. In this respect a crossing-recrossing correction
deserves special attention. In section III, the energy estimation is
considered, with special attention to the flux-term coming from the
absorption of walkers at the boundary. In the last section the results for a
set of oscillator models is given, followed by a discussion and conclusions.

\section{Free Diffusion With A Boundary}

For $N$ particles moving in a $3N$-dimensional configuration space,
permutation symmetry can be taken into account by ordering linearly the
coordinates of the particles along the three basic directions, introducing
in this way a domain $D_N^{\,3}$. Depending on the configuration the
antisymmetric representation can be realized by several combinations of
boundary conditions: the antisymmetric representation is {\it irreducible}
with respect to the permutation of $N$ particles but is {\it reducible} with
respect to the permutation of the $3\times N$ coordinates. The symmetry of
the model without interaction or with only harmonic interactions along the
principal axes is higher than that of a many body system with realistic
one-body and two-body potentials. A decomposition into irreducible
representations of the permutation of coordinates leads to four
combinations, labeled by a quantum number $\ell =0,1,2,3$ referring to the
behavior of the coordinates at the boundary $\partial D_N^{\,u}$, $u=x,y,z$.
We follow here the conventions introduced in \cite{BDL96}. The case $\ell
\!=\!0$ for the antisymmetric combination e.g. means that coordinates behave
fermion-like in the three directions, while $\ell \!=\!2$ for instance means
that the coordinates behave fermion-like in the $y$ direction and boson-like
in the two other directions. Using this reduction by symmetry and the
appropriate boundary conditions, the representative domain that is specified
can not be left by the walkers anymore. On this domain the wave function of
the model does not change sign, leading to a sign-problem-free simulation of
that antisymmetric state. It is important first to analyze the algorithm for
moves in one direction only. This is done for fermion-like as well as for
boson-like behavior. The moves in a general $3D$-direction are combinations
of both.

\subsection{Reflection and bosons on a line}

For $N$ bosons moving on a line, the domain $D_N$, that we consider here, is
linearly ordered by $x_i>x_{i-1}$ and the boundary of the domain $\partial
D_N$ is reached if one of the following equations $x_i=x_{i-1}$ is satisfied
for $i=2\ldots N$. A sample path for $N$ walkers starting in the domain is
denoted by $X(\tau )$ and the increments of this $N$-dimensional random
walker or Brownian motion are independent and obtained as usual. This sample
path can be used to construct one with reflection at the boundary $\partial
D_N$ by the following procedure. Consider the path of one component $i$ with
a reflecting boundary at the origin. In this case those parts of $X_i\left(
\tau \right) $ which are {\it outside} the domain before the time $\tau $
are reflected to the {\it inside} by adding an amount $L_i\left( \tau
\right) $, determined by the most negative value on the trajectory in the
time interval $\left[ 0,\tau \right] $%
\begin{equation}
L_i\left( \tau \right) =-\inf \left\{ X_i\left( \tau ^{\prime }\right)
,0;\tau ^{\prime }\leq \tau \right\} .
\end{equation}
The sample path of the diffusion or random walk with reflection is given by
the Skohorod construction: 
\begin{equation}
\hat{X_i}\left( \tau \right) =X_i\left( \tau \right) +L_i\left( \tau \right)
.  \label{sko}
\end{equation}
By definition, $L_i\left( \tau \right) $ is an non-decreasing random
function on the event space, and it increases only when $\hat{X}_i\left(
t\right) $ reaches the boundary, i.e. in this example if $\hat{X}_i\left(
\tau \right) =0$. This mechanism is illustrated in fig. 1, in which the
reflection of a Brownian sample path $X_i\left( \tau \right) $ occurs by the
combination of $X_i\left( \tau \right) $ with $L_i\left( \tau \right) .$
(For vividness but without consequences for the principle in mind, the
sample path has been depicted as a smooth function.) With the $N$%
-dimensional generalization of the construction, it can be shown that the
transition probability satisfies the Neumann boundary condition for the
backward equation. It should be noted that if a diffusion with drift is
considered - the forward and backward equations are different - the boundary
conditions will become dependent on the drift \cite{BathWay}.

In order to see how the distribution arises in a simulation using
prescription (\ref{sko}) for boson diffusion, the example of two identical
bosons on a line is instructive. A representative outcome of the numerically
simulated free-particle boson diffusion is compared in fig. 2 with the
rigorous densitiy, i.e. a permanent with the appropriate transition
probabilities $\rho \left( x_i^{\prime },x_j;\Delta \tau \right) $ as
entries, where $x_i^{\prime }$ and $x_j\ $(with $i,j=1,\ldots N)$ are
coordinates at time $\tau $ and at time $\tau +\Delta \tau $ respectively.
First, an ensemble of walkers was created at a fixed position inside the
state space $D_2$. The initial position is indicated by a cross, whereas the
state space $D_2$ with $x_1>x_2$ is situated right below the solid black
lines $x_1=x_2$. The walkers evolved for three atomic times and by
projection of the walker coordinates onto the $\left( x_1,x_2\right) $-plane
a snapshot of the simulated densitiy has been obtained. The numerical
density is in good agreement with the rigorous one, also shown in the figure.

\subsection{Absorption and fermions on a line}

For fermions on a line on the same domain $D_N$ but with absorbing boundary
conditions, the sampling technique has to generate a determinant with as
elements the $\rho \left( x_i^{\prime },x_j;\Delta \tau \right) $. The
construction of a sample path with absorption from a Brownian motion or a
random walk $X(\tau )$ proceeds with the introduction of a first-passage
time $\tau _{\partial D}$: 
\begin{equation}
\tau _{\partial D}=\inf \left\{ \tau \geq 0;X\left( \tau \right) \in
\partial D_n\right\} .
\end{equation}
The new stochastic process $\widetilde{X}(\tau )$ is then by definition: 
\begin{equation}
\widetilde{X}(\tau )=\left\{ 
\begin{array}{lll}
X(\tau ) & \text{ if } & \tau <\tau _{\partial D}, \\ 
X_{\tau _{\partial D}} & \text{ if } & \tau \geq \ \tau _{\partial D}.
\end{array}
\right.  \label{const}
\end{equation}
It should be noted that each sample path $X(\tau )$ has its own
first-passage time; therefore this time is a random variable. The transition
probability for the process $\widetilde{X}(\tau )$ satisfies the Dirichlet
boundary conditions. As argued in \cite{BDL94}, this procedure simulates the
Slater determinant describing motion on a line.

In order to see how the distribution arise using prescription (\ref{const})
for fermion diffusion, the simulation of two identical fermions on a line is
shown in fig. 3. Also in this case, the numerical density has been found to
be in good agreement with the analytically rigorous one.

\subsection{Bosons and Fermions in $1D$}

The figures 2-3 vividly recover typical characteristics attributing to the
time evolution of free bosons and fermions. As has been shown in a previous
paper \cite{BDL95}, the state-space boundary $\partial D_2$ acts attractive
for the two bosons. Their modal trajectory, i.e. the trajectory with the
highest density, rapidly approaches the boundary and stays there for the
rest of the evolution time \cite{BDL95}. Fermion evolution on the contrary
looks as if the two fermions would be repelled by the boundary; their modal
distance monotonously increases with Euclidean time. These properties are in
contrast to the evolution of distinguishable particles, which is not
restricted to the state space $D_2$ (see fig.4). The walkers spread out
isotropically with rigid modal distance, what nicely illustrates the lack of
free-particle interactions due to quantum statistics.

\subsection{Absorption and reflection in more than $1D$}

In order to illustrate the construction of an antisymmetric state for two
particles using the appropriate processes for diffusion of their
coordinates, the rate that the walkers get absorbed at the boundary of the
state space is a characteristic of the many-body diffusion algorithm. In
three spatial dimensions and for an antisymmetric state, the four fermion
processes \{f,b,b\}, \{b,f,b\}, \{b,b,f\} and \{f,f,f\} introduced in \cite
{BDL96} generally reveal different absorption rates. Fig. 5 gives a typical
example for the (Euclidean-)time-dependence of the relative number of
walkers residing in the state space. Providing infinitely large samples,
this numerical quantity converges to the conditional probability $P_{%
\overline{x}}(\tau _{\partial D}>\tau )$. Due to the presence of absorbing
boundaries in each direction, the \{f,f,f\}-process decays faster than any
of the other three processes \{f,b,b\}, \{b,f,b\} and \{b,b,f\}. As can be
clearly seen in fig. 5, the processes decay with a different rate; the
reason for this is that the distribution of the first-passage time $\tau
_{\partial D}$ depends on the initial position. For large times, $\tau \to
\infty $, they decay like $\tau ^{-1/2}$ proportional to the relative
fermion-like coordinate of the initial position. Again, considering the
state space as the domain of interest, the evolution of the {\it exit} time
for distinguishable particles reveals a totally different decay in time,
because nothing prevents them from reentering the domain after having left
it before.

\section{Sampling the ground state}

The sample paths inside the domain are made using the traditional techniques
for Brownian motion or Bernoulli walks. In addition, computational
efficiency calls for the incorporation of the Feynman-Kac functional using
the technique of branching and killing. This is consistent with the
underlying formalism of many-body diffusion provided the system does not
require walker transitions from an $\ell $-state to a (different) $\ell
^{\,^{\prime }}$-state, thereby changing its boundary conditions. \ The fact
that such transitions are possible in a general potential has been
overlooked in \cite{BDL96} (see \cite{BDLer97}). However, it remains of
interest to simulate a class of antisymmetric states sign-problem free with
the proposed algorithm; in particular, in view of algorithms for more
advanced models taking into account the additional transitions from $\ell $
to $\ell ^{\prime }$.

\subsection{Estimation of the first-passage time}

Although the consequences of permutation symmetry are rigorously
incorporated in both the formalism of many-body diffusion and the algorithm
described so far, numerical practice reveals that the simulation of the
imaginary-time Schr\"{o}dinger equation along these lines with {\sl discrete}
time steps introduces systematic inaccuracies due to the possibility of
crossing and recrossing the boundary during that finite time step. To
elucidate this point, we focus on the evolution of free particles, for the
effect is caused by the underlying kinetics, the potential here being
irrelevant. Consider two identical fermions on a line, with their
coordinates in the configuration space denoted as $x_1$ and $x_2$ (see fig.
6). The absorbing boundary is given by: 
\begin{equation}
\partial D_2=\left\{ x_1=x_2\right\} 
\end{equation}

Assume that a discrete time scale $\Delta \tau $ and, consistently according
to the Einstein relation, a spatial grid with spacings $\Delta x=\sqrt{%
\Delta \tau }$ has been chosen, as illustrated by fig.5a. During a time step 
$\Delta \tau $, a walker initially located at $A$ could move a distance $%
\Delta x$ into each direction, so to reach one of the points $C$, $C^{\prime
}$, $C^{\prime \prime }$ or $C^{\prime \prime \prime }$. Let it eventually
advance to $C$. The only possibility to perform that step on the broad grid
is thus the straight move, as indicated by the straight solid line $AC$ in
fig.5a. Refinement of the grid (see fig.5b) to spacings $\Delta x/2$, and
correspondingly time steps $\Delta \tau /4$, indicates that not all possible
paths from $A$ to $C$ provide non-zero contributions. During the four time
steps $\Delta \tau /4$, the diffuser could either have taken admissible
paths $AC$ like the dashed one, or -- on the opposite -- could meanwhile
have left the domain like indicated by the solid path $AC$. In the latter
case, the move would be invalid and the diffuser should be absorbed. But the
broad grid in fig.5a does principally not allow to distinguish between valid
and invalid paths $AC$. In other words, working on any (artificial) discrete
grid, any move the initial and final positions of which are situated inside
the domain automatically includes a set of invalid paths. This gives rise to
systematic errors in the evolution.

\subsection{The crossing/recrossing correction}

To correct for these errors in the evolution, each trial displacement of a
diffuser must be assigned a ratio of validity, which directly corresponds to
the ratio of possible valid and invalid paths to perform the move in
continuous space. The correction has to do with the crossing and recrossing
at the boundary in a time lapse smaller than the time increment used in the
simulation. It may be illustrated by the following arguments.

A probabilistic argument based on the reflection principle can be given as
follows. Suppose that a walker, situated at $A$, evolves to $C$ (see fig.7).
It might either have taken a completely valid path (solid line) or have hit
the boundary at e.g. $B$ (dashed line $ABC$ in fig. 7). Because free
distinguishable particle diffusion is isotropic, the probability $P\left[
X_{\tau +\Delta \tau }=C\mid X_\tau =A,\ X_\nu =B\ \right] $ to reach $C$
over an arbitrary point $B$ on $\partial D_2$ with $\nu \in \left[ \tau
,\tau +\Delta \tau \right] $, equals the probability $P[X_{\tau +\Delta \tau
}=C^{^{\prime }}\mid X_\tau =A,\ X_\upsilon =B]$ to end up in the image $%
C^{\prime }$. Clearly, the total probability for the move from $A$ to $C$ is
the sum of the two conditional probabilities to move either from $A$ to $C$
without crossing $B$ or to move from $A$ to $C^{\prime }$ with crossing $B$.
Generalizing this argumentation to all points $B$ on the boundary, the move
from $A$ to $C$ has to take into account two non-intersecting subsets of
paths, i.e. 
\begin{equation}
P\left[ X_{\tau +\Delta \tau }=C\mid X_\tau =A\right] =P_\partial \left[
X_{\tau +\Delta \tau }=C\mid X_\tau =A\right] +P_{\widetilde{\partial }%
}\left[ X_{\tau +\Delta \tau }=C\mid X_\tau =A\right] ,
\end{equation}
where the index $\partial $ means that the path crosses the boundary, while $%
\widetilde{\partial }$ indicates that $C$ is reached without boundary
crossing/recrossing. The (conditional) probability $P_\partial \left[
X_{\tau +\Delta \tau }=C\mid X_\tau =A\right] $ equals $P_\partial \left[
X_{\tau +\Delta \tau }=C^{^{\prime }}\mid X_\tau =A\right] $, and since any
move from $A$ to $C^{\prime }$ crosses the boundary one obtains 
\begin{equation}
P\left[ X_{\tau +\Delta \tau }=C\mid X_\tau =A\right] =P[X_{\tau +\Delta
\tau }=C^{^{\prime }}\mid X_\tau =A]+P_{\widetilde{\partial }}\left[ X_{\tau
+\Delta \tau }=C\mid X_\tau =A\right] .
\end{equation}
The crossing/recrossing correction can then be identified with 
\begin{equation}
P_{\widetilde{\partial }}\left[ X_{\tau +\Delta \tau }=C\mid X_\tau
=A\right] =P\left[ X_{\tau +\Delta \tau }=C\mid X_\tau =A\right] \left( 1-%
\frac{P[X_{\tau +\Delta \tau }=C^{^{\prime }}\mid X_\tau =A]}{P\left[
X_{\tau +\Delta \tau }=C\mid X_\tau =A\right] }\right) .
\end{equation}
Hence, the normalized probability for moves from A to C with boundary
crossing/recrossing arises from the ratio 
\begin{equation}
R(A\!\to \!C)=\frac{P[X_{\tau +\Delta \tau }=C^{^{\prime }}\mid X_\tau =A]}{%
P\left[ X_{\tau +\Delta \tau }=C\mid X_\tau =A\right] }=\exp \!\left\{ -%
\frac{(x_1^{\prime }-x_2^{\prime })(x_1-x_2)}{\Delta \tau }\right\} \ \ \ .
\label{suppl}
\end{equation}
A physical argument leading to the same correction is based on the method of
the images implying that the evolution of two walkers without interaction
but with absorbing boundary conditions is given by the determinant of the
free propagators. Factoring out the diagonal element one obtains the same
expression. In practice, the evolution procedure must be extended with the
supplementary rejection ratio (\ref{suppl}): even diffusers having made a
move to a final position which is clearly situated inside the domain may
eventually be judged invalid.

As the many-body diffusion formalism traces back the evolution of many-body
systems of arbitrary spatial dimension to adapted one-dimensional diffusion
processes, the argument described above is readily generalized to any
dimensions. The extension to $N$ particles with e.g. ordering along the
x-axis corresponds to the construction of $N-1$ significant boundary
conditions. The probability to cross the set of boundaries during the move
from A to C, both points elements of the domain, is complementary to the
probability not to hit any of the $N\!-\!1$ boundaries. Hence, a completely
valid path occurs according to the product of elemental probabilities of the
form $1\!-\!R(A\!\to \!C)$. The implementation of these extended
probabilities eliminates the systematic errors originating from
finite-time-step considerations in the presence of absorbing boundaries.
Strictly speaking, the systematic errors specific for our numerical
description of many-fermion systems compares with systematic inaccuracy
attributing to the simulation of respective distinguishable-particle systems.

It must be stressed that the arguments concerning boundary
crossing/recrossing effects are reminiscent of the method of image charges,
which have been transferred to probability theory a long time ago \cite
{FELLE}. The method is based on the mathematical formalism of diffusion as
distinct from other up-to-date quantum Monte Carlo techniques. Quite in
contrast, the same geometrical argument has been employed by Anderson \cite
{AND76} to formulate the behavior at boundaries fixed by the nodal structure
of a trial function and by \cite{Trif96} to introduce the so called
Pauli-potential.

\section{Estimation Of Ground-State Energies}

Most often propagator-based methods will be used to estimate the energy of
the system under consideration. In this case, one can use the exponential
decay of the free energy, or one may devise separate estimators for the
components of the internal energy. Propagator-based methods use implicitly
paths with known begin and end points. The key path integral goes over all
paths starting and ending in the same point of the phase space. In the
algorithm explained in the preceding section only a construction of sample
paths with a known starting point is given. A generalization of the
construction, keeping the increments independent and imposing an end point
but taking the boundary into account, is to our knowledge not documented in
the mathematical literature. Without the boundary condition the construction
is known as the Brownian bridge; physically it means that one takes the
classical action as the starting point to incorporate quantum fluctuations.
Without the generalization of the Brownian bridge another estimation scheme
for the ground-state energy has to be used.

\subsection{The kinetic contribution}

To the best of our knowledge, it was Anderson \cite{AND75} who first
reported ground-state energy estimates based on the a priori assumption of
the exponential ground-state-energy dependence of the time decay of the
propagator. For distinguishable-particle systems, the ground-state energy
can be evaluated by `stochastic' averages over the potential $V(\overline{r}%
) $ 
\begin{equation}
E_0\,=\,\lim_{\tau \to \infty }E(\tau )\;=\;\lim_{\tau \to \infty }\frac{%
\int d\overline{r}\;\Psi (\overline{r},\tau )\,V(\overline{r})}{\int d%
\overline{r}\;\Psi (\overline{r},\tau )}=\langle V\rangle ^s.
\label{energie}
\end{equation}
The `stochastic estimates' $E(\tau )$ are based on the stochastic
probability density 
\begin{equation}
\Psi (\overline{r},\tau )={\Bbb E}_R\left[ \exp \left( -\int\limits_0^\tau
V\left( R\left( \widetilde{\tau }\right) \right) d\widetilde{\tau }\right)
\right] \,.  \label{wav}
\end{equation}

For sufficiently large Euclidean time $\tau $, $\Psi (\overline{r},\tau )$
mimics the ground-state wave function and, accordingly, the estimates $%
E(\tau )$ reflect the ground-state energy $E_0$.

The expectation in (\ref{wav}) goes over all sample paths starting at $%
\overline{r}$ and ending wherever in the configuration space with $\left\{
X\left( \tau \right) ;\tau \geq 0\right\} $ a Brownian motion or a random
walk. This type of functional integration leading to a wave function is
based on the path-integral representation of the wave operator \cite{Roep}.
Once a procedure for the wave function is found, it is obvious that the
potential average (\ref{energie}) is in direct contrast to the
quantum-mechanical expectation value of the potential, and both potential
averages must be clearly distinguished.

Antisymmetry, as specific for the ground-state wave function of identical
fermions, causes the breakdown of the potential average (\ref{energie}) as
an estimate of the ground-state energy in the configuration space. Both the
nominator and the denominator in the estimator tend to zero: the projection
on the antisymmetric representation of the simulated wave function in the
asymptotic limit of large Euclidean time $\tau $ is not well-defined
anymore. During the thermalization of the system, the statistical error
attached to the numerical estimates rapidly overwhelms the signal.

Replacing the Brownian motion or random walk over the configuration space by
the many-body diffusion process as discussed in the preceding section, the
formulation of energy estimates adapted to the state space $D_N$ with its
boundary conditions avoids the mentioned decay of the signal-to-noise ratio.
The implementation of the many-body diffusion process goes hand in hand with
a loss of walkers at the absorbing boundaries of the state space. From the
algorithmic point of view, this loss is caused by the underlying diffusion
kinetics and contains essential information on the lowest energy of that
antisymmetric state. In terms of the wave function on the domain, (\ref
{energie}) has to be replaced by: 
\begin{equation}
E_0\,=\,\lim_{\tau \to \infty }E_\tau \;=\;\lim_{\tau \to \infty }-\frac{%
\frac{\hbar ^2}{2m}\int_{\partial D_N}d\overline{r}\;\overline{\nabla }\Psi
_F(\overline{r},\tau )}{\int_{D_N}d\overline{r}\;\Psi _F(\overline{r},\tau )}%
+\frac{\int_{D_N}d\overline{r}\;\Psi _F(\overline{r},\tau )\,V(\overline{r})%
}{\int_{D_N}d\overline{r}\;\Psi _F(\overline{r},\tau )}\,=\,\langle
\,j\,\rangle ^s+\langle \,V\,\rangle ^s\ \ \ ,  \label{energ}
\end{equation}
where $\Psi _F(\overline{r})=\Psi _F(\overline{r},\infty )$ denotes the
asymptotic probability density on the state space $D_N$ with the boundaries $%
\partial D_N$: 
\begin{equation}
\Psi _F(\overline{r},\tau )={\Bbb E}_{R_F}\left[ \exp \left(
-\int\limits_0^\tau V\left( R_F\left( \widetilde{\tau }\right) \right) d%
\widetilde{\tau }\right) \right] ,
\end{equation}
where $\left\{ X_F\left( \tau \right) ;\tau \geqslant 0\right\} $ is a many
body diffusion or many body random walk.

In contrast to the distinguishable-particle estimator (\ref{energie}), eq. (%
\ref{energ}) comprises additional kinetic contributions $\langle j\rangle ^s$%
. In the case of fermions, $\langle j\rangle ^s$ can be interpreted as the
relative flow per time step through the absorbing boundaries of the state
space, this expectation value depends on the distribution of the
first-passage time $\tau _{\partial D}$. In the standard treatment of
distinguishable particles, $\overline{\nabla }\Psi (\overline{r})$ at the
corresponding (infinitely remote) boundaries is zero. Evidence that the
gradient of diffusive movement at reflecting boundaries vanishes has been
given in \cite{BDL96}. For a diffusion or a random walk of the
\{f,b,b\}-type, for example, the kinetics reveals non-zero gradients at the
absorbing boundaries in the $x$-direction. Consequently, surface terms $%
\langle j\rangle ^s$ are of essential importance in our treatment.

\subsection{The model system}

The algorithm introduced in the previous section has been used to calculate
the lowest energy of the states specified by a given symmetry for an
oscillator model with the following Hamiltonian: 
\begin{equation}
H=-\sum\limits_{j=1}^N\frac{\hbar ^2}{2m}\Delta _j+\sum\limits_{j=1}^N\frac{%
m\Omega ^2}2\vec{r}_j^{\,2}-\sum\limits_{i,j=1}^N\frac{m\omega ^2}2(\vec{r}%
_i-\vec{r}_j)^2.  \label{oscsys}
\end{equation}
The repulsive interparticle interaction is introduced to mimic many-body
interaction. Because (\ref{oscsys}) can be diagonalized, all the energies of
the system are known analytically. The diagonalization, the thermodynamic
properties and the response properties of this model have been discussed
elsewhere \cite{BDLPRE97a,BDLPRE97b}. The ground states of (\ref{oscsys})
for identical particles with a specific symmetry (i.e. specifying the
boundary conditions on the domain $D_N^3$) are excited states of (\ref
{oscsys}), considered as a system of distinguishable particles. The purpose
of showing that the algorithm is workable has led us to choose the worst
cases: the particles are arranged in such a way that the ranking according
to all their coordinates is maintained during the simulation, and fermion
diffusion is chosen along one direction, while boson diffusion is chosen for
the other directions. The simulation has been done for several values of $%
\Omega $ and $\omega $ indicated in Table 1. $N=N_{\uparrow }+N_{\downarrow
} $ indicates the number of particles in the system. The $N_{\uparrow }$
spin-up particles and the $N_{\downarrow }$ spin-down particles are mutually
indistinguishable, but the $N_{\uparrow }$ spin-up particles are
distinguishable from the $N_{\downarrow }$ spin-down particles. $D$ is the
spatial dimension of the evolution equation for each particle. $E$ is the
energy obtained from the simulation, $\sigma $ is the standard deviation, $%
E_{\text{exact}}$ is the analytically obtained energy value of the excited
state of the distinguishable oscillator model corresponding with the
symmetry as explained above. In order to have an idea about the relevance of
the flux term and of the boundary conditions, the exchange energy $E_{\text{%
exc}}$ is the difference between $E$ and the ground state energy of the
system of distinguishable particles. Time step periods of 10$^{-3}$ atomic
units proved suitable for the parameter values listed in table 1.

The present results confirm the accurate implementation of quantum
statistics for the model of indistinguishable interacting harmonic
oscillators. In all the considered cases, the many-body diffusion algorithm
has achieved energy estimates, which are in good agreement with the
analytical rigorous results. We stress, that our results do not
significantly depend on the choice of the initial densities. In particular,
the support of sophisticated trial functions has been avoided in the present
paper, for the emphasis is on the investigation of the algorithm's general
capability to describe indistinguishable particles.

\section{Discussion and Conclusions}

It should be emphasized that in all the simulations that we have reported
here, use has been made of the special symmetry property of the model under
study. Its Hamiltonian is invariant under permutations of the Cartesian
components of the coordinates. For motion in 3D, this invariance allows to
define four independent partial processes, each with a well-defined boundary
condition. Introducing a number of walkers, initially randomly divided over
the four types of processes, their absorption determines how many walkers
survive. The absorption together with the reflection at the boundary $%
\partial D_N^{\,3}$ determines the distribution in $D_N^{\,3}$. This has
been illustrated for a few low-dimensional examples. Given that the
distribution is understood, an estimator for the ground-state energy is
obtained. This estimator takes the out-flow of the walkers out of the domain
into account as a crucial ingredient in the ground-state energy. In order to
study the statistical errors two common simulation methods have been
considered; both the inclusion of Bernoulli walks or diffusion leads to
analogous results if a suitable time step is chosen (that also determines
the unit of distance by the fundamental diffusion constant $\hbar /2m$
through the Einstein relation) and if the crossing-recrossing corrections
for absorbing boundaries are applied. The ground-state energy estimation is
satisfactory, especially taking into consideration that these estimates are
obtained without invoking importance sampling. Furthermore, the estimation
is sign-problem-free.

A discussion of the present results is not an easy matter when placed
against the large number of publications, partially dealing with this
problem. In the long run, it will be the ability to treat realistic models
with an extension of this method, that will determine its usefulness.
Without further discussion and without being exhaustive the following
references deal with such models \cite{TAK84,NEW92,HAL92,CHA93}.

In summary, in this paper a sign-problem-free simulation of a specific
interacting fermion system is presented. The simulated system has harmonic
interactions, and the algorithm as well as the theory leading to this
algorithm make explicit use of the harmonic character of the interactions.
Apart from this limitation, it is shown that the algorithm based on the
absorption and reflection at a boundary of a Brownian motion or a Bernoulli
walk reproduces the correct distributions. It gives also good energy
estimates when crossing and recrossing correction on the first-exit time are
taken into account. In view of the fact that no importance sampling has been
used an acceptable standard deviation has been obtained.

\acknowledgments

Work supported by the Interuniversity Poles of Attraction Programme -
Belgian State, Prime Minister's Office - Federal Office for Scientific,
Technical and Cultural Affaires. We acknowledge support by the BOF NOI 1997
projects of the University of Antwerpen, the NFWO project WO.073.94N
(Wetenschappelijke Onderzoeksgemeenschap, Scientific Research Community of
the NFWO on ''Low-Dimensional Systems'') and the European Community Program
Human Capital Mobility through Contract No. CHRX-CT93-0124. Part of this
work has been performed in the framework of the projects G.0287.95 and the
supercomputer project 1.5.729.94N of the FWO. F.L. is very grateful to B.
Gerlach for partial financial support during the development of this work.

\newpage \label{0table}

\begin{table}[tbp] \centering%
\begin{tabular}{llllllllll}
\hline\hline
$\Omega $ & $\omega $ & $N$ & $N_{\uparrow }$ & $N_{\downarrow }$ & $D$ & $E$
& $\sigma $ & $E_{\text{exact}}$ & $E_{\text{exc}}$ \\ \hline\hline
2 & 1 & \multicolumn{1}{r}{2} & \multicolumn{1}{r}{2} & \multicolumn{1}{r}{0}
& 1 & 3.1215 & 0.0010 & 3.1213 & 1.4142 \\ 
2 & 1 & \multicolumn{1}{r}{2} & \multicolumn{1}{r}{2} & \multicolumn{1}{r}{0}
& 3 & 6.5356 & 0.0013 & 6.5355 & 1.4142 \\ 
2 & 1 & \multicolumn{1}{r}{3} & \multicolumn{1}{r}{3} & \multicolumn{1}{r}{0}
& 3 & 9.0091 & 0.0051 & 9.0000 & 3.0000 \\ 
3 & 1 & \multicolumn{1}{r}{4} & \multicolumn{1}{r}{2} & \multicolumn{1}{r}{2}
& 3 & 19.039 & 0.0074 & 19.034 & 4.4721 \\ 
3 & 1 & \multicolumn{1}{r}{6} & \multicolumn{1}{r}{2} & \multicolumn{1}{r}{4}
& 3 & 29.594 & 0.0185 & 29.615 & 12.124 \\ 
3 & 1 & \multicolumn{1}{r}{8} & \multicolumn{1}{r}{7} & \multicolumn{1}{r}{1}
& 3 & 35.991 & 0.0209 & 36.000 & 21.000 \\ 
4 & 1 & \multicolumn{1}{r}{10} & \multicolumn{1}{r}{10} & \multicolumn{1}{r}{
0} & 3 & 149.27 & 0.0989 & 149.30 & 110.23 \\ 
4 & 1 & \multicolumn{1}{r}{10} & \multicolumn{1}{r}{5} & \multicolumn{1}{r}{5
} & 3 & 88.118 & 0.0472 & 88.058 & 48.990 \\ 
5 & 1 & \multicolumn{1}{r}{20} & \multicolumn{1}{r}{10} & \multicolumn{1}{r}{
10} & 3 & 272.52 & 0.2795 & 272.47 & 201.25
\end{tabular}
\caption{Comparison of eigenenergies (a.u.) gained by the MBDA, E, with
results from rigorous analytical calculations.\label{key}}%
\end{table}%

\begin{center}
{\sc Figure Captions}
\end{center}

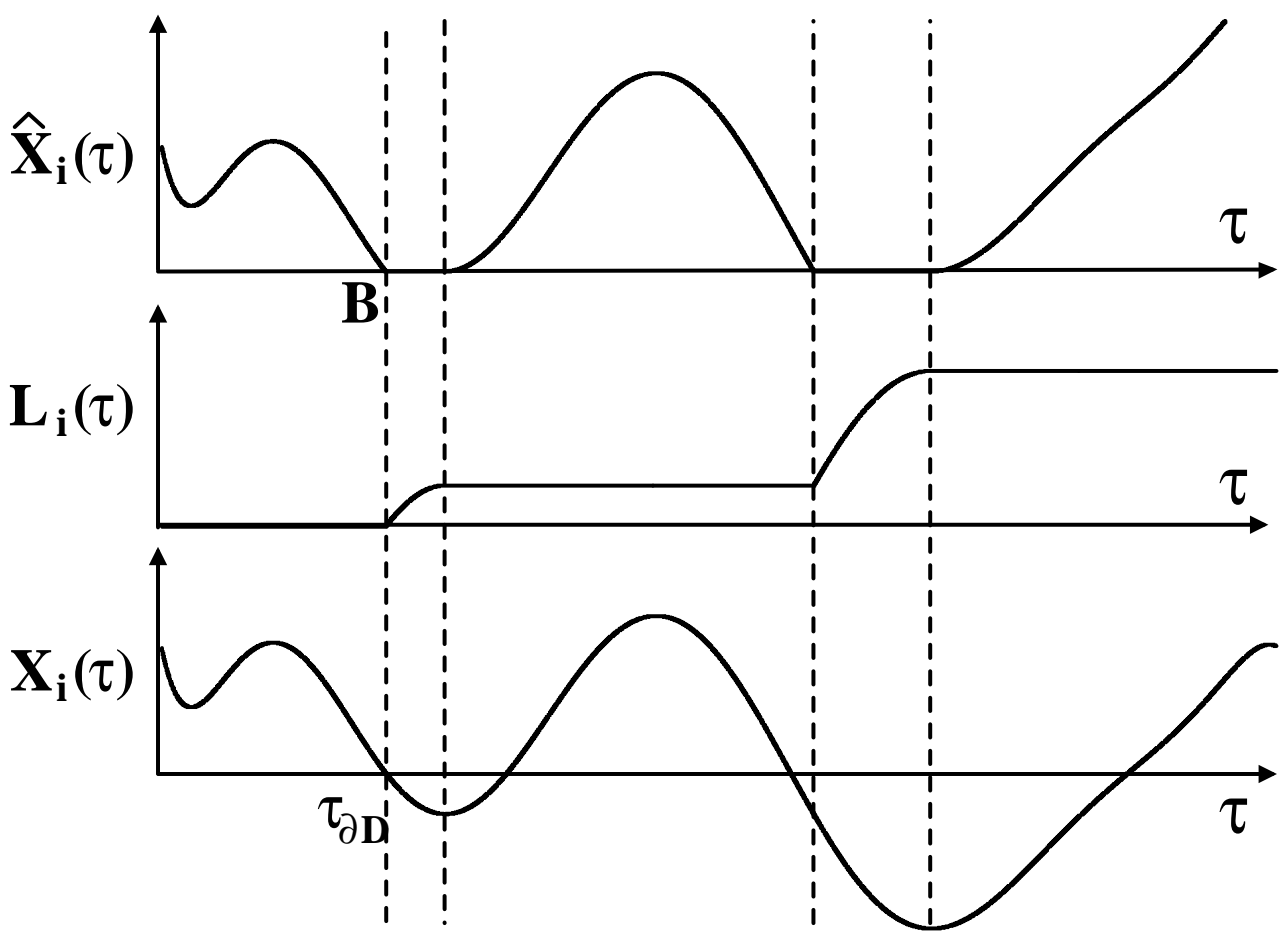
{\bf Fig. 1: }Illustration of the Skohorod construction for a Brownian
trajectory $X_i(\tau )$. $\widehat{X}_i(\tau )$ refers to the boson-like
random process in the presence of a reflecting boundary ($\tau $-axis). The
corresponding fermion-like process $\widetilde{X}_i(\tau )$ is identical
with $X_i(\tau )$ till the point $B$ whereupon it stays on the boundary
propagating with zero weight.

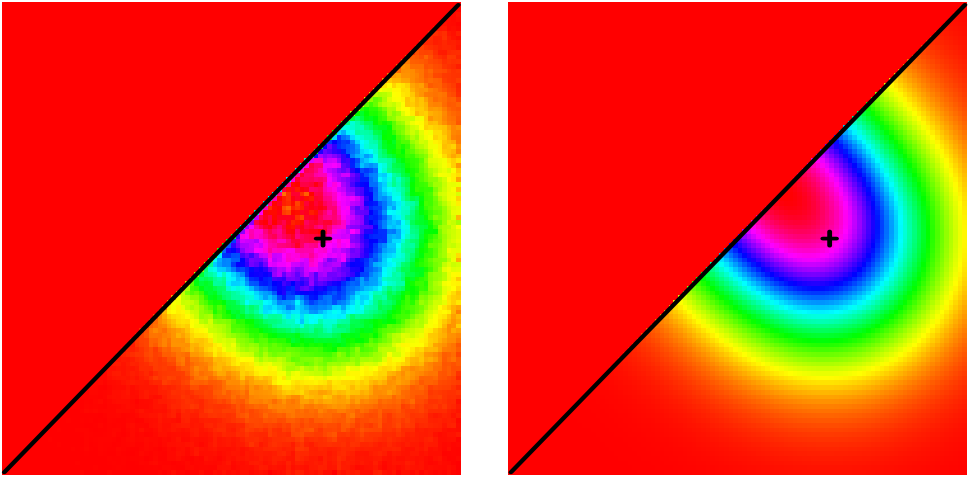
{\bf Fig. 2.: }Comparison of the rigorous (right hand side) and numerically
sampled (left hand side) probability densities of two one-dimensional
free-diffusive identical bosons, having started from an initial point
denoted by the crosses, after an evolution of three atomic time units. The
solid line indicates the state space boundary $x_1=x_2$.

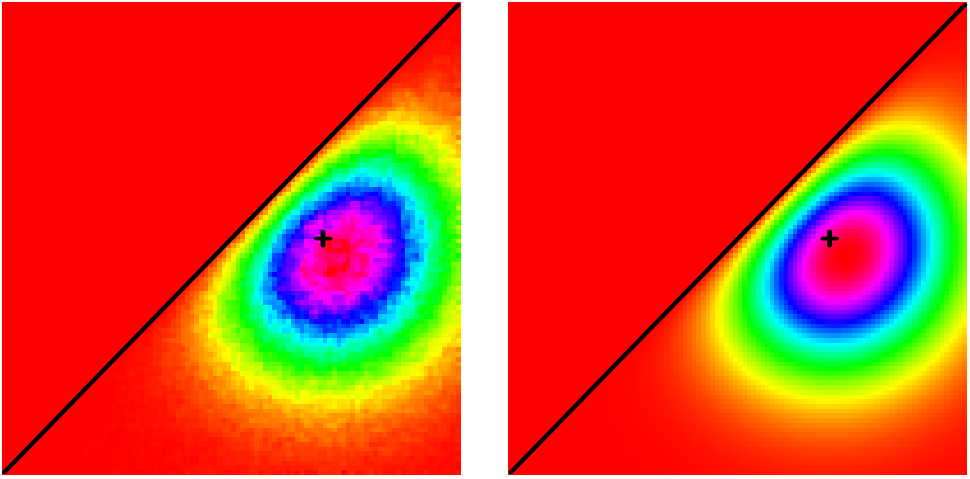
{\bf Fig. 3: }The same as fig. 2 for fermions.

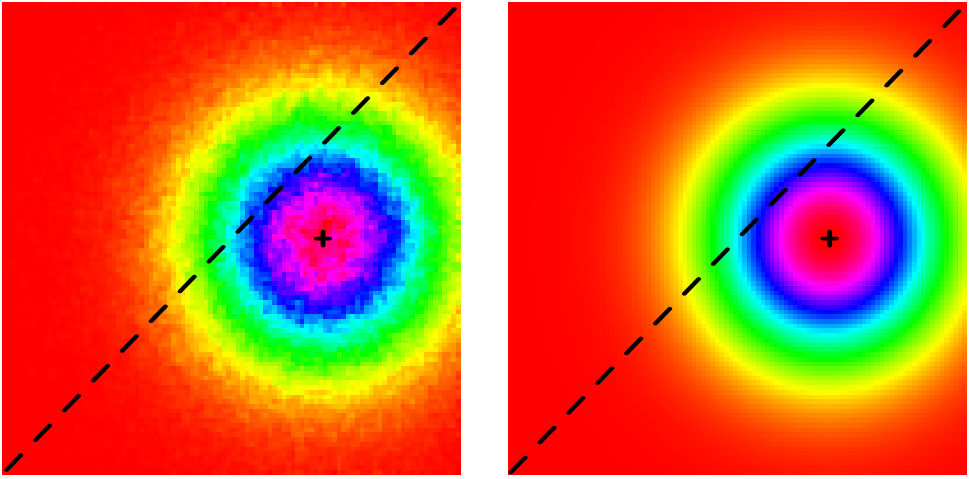
{\bf Fig. 4: }The same as fig. 2 for distinguishable particles. The dashed
line indicates the hypersurface $x_1=x_2$.

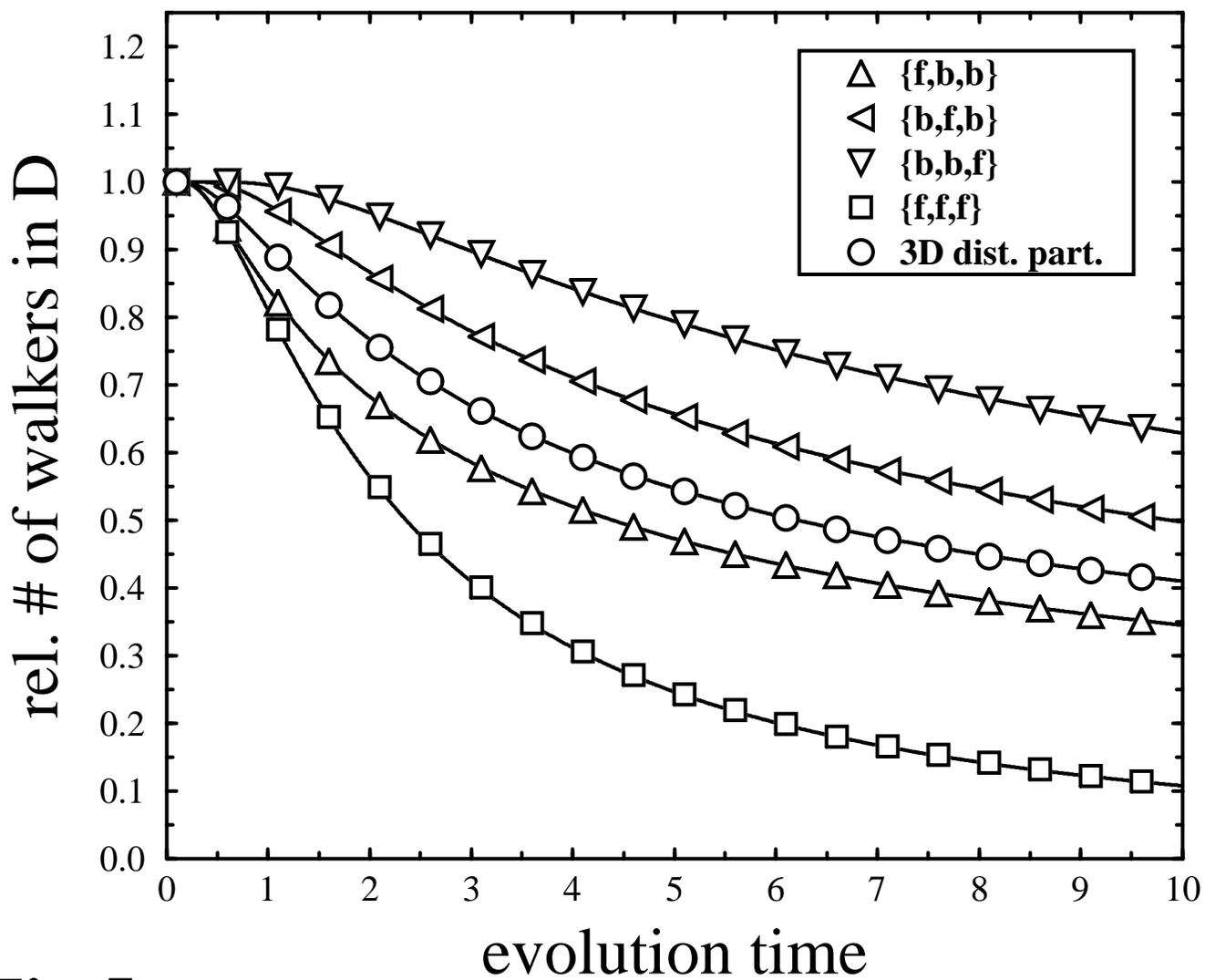
{\bf Fig. 5: }An exemplary comparison of the numerically obtained relative
number of walkers inside the state space with the rigorous results indicated
by the solid lines, as a function of evolution time for both the
subprocesses modelling the many-body diffusion process of two
three-dimensional identical fermions and three-dimensional
free-distinguishable-particle diffusion.

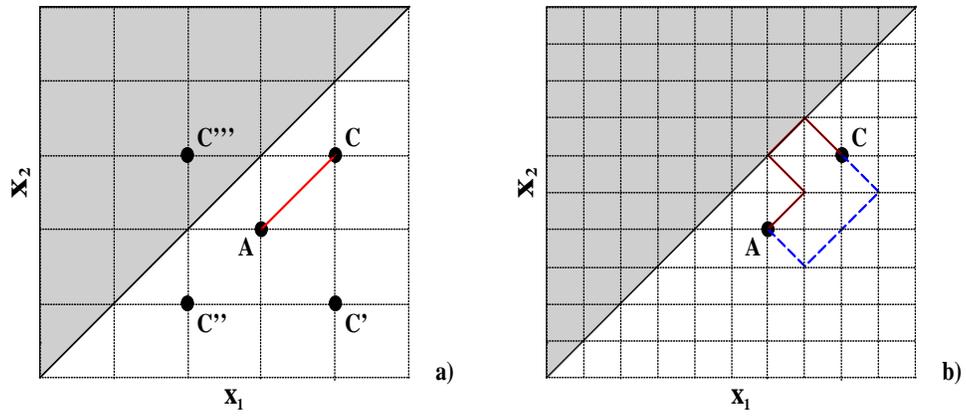
{\bf Fig. 6: }The definition of finite time steps and corresponding spatial
grids induces shortcomings with respect to the accurate assignment of
validity flags. The refinement of the coarse grid in a), indicates [see b)]
that a considerable rate of possible paths from $A$ to $C$ enters prohibited
(grey) regions.

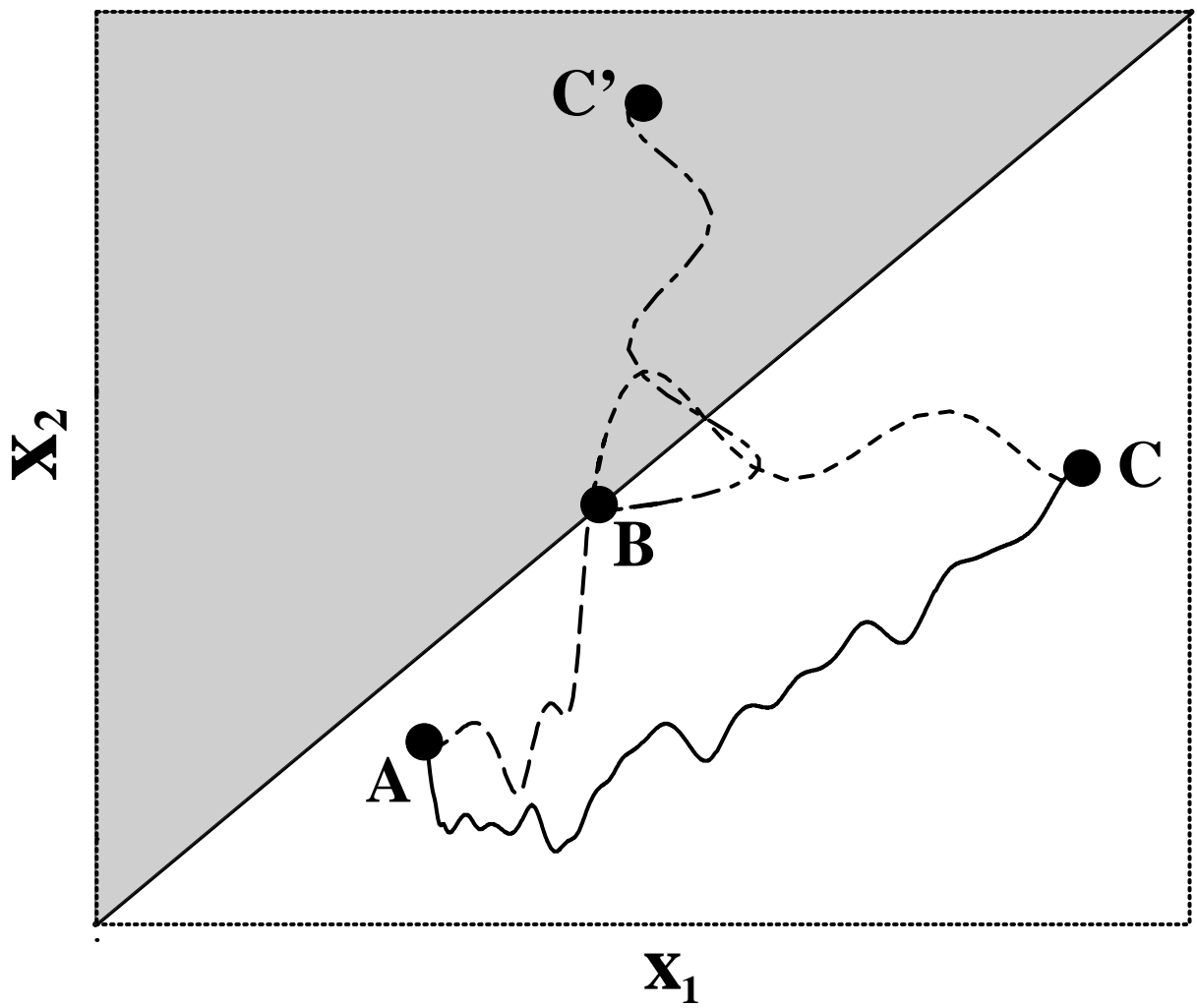
{\bf Fig. 7: }Diffusers having propagated from $A$ to $C$ may have followed
a valid path (solid line) as well as an invalid one (dashed line $ABC$)
which passed into the forbidden region (in grey) outside the domain.

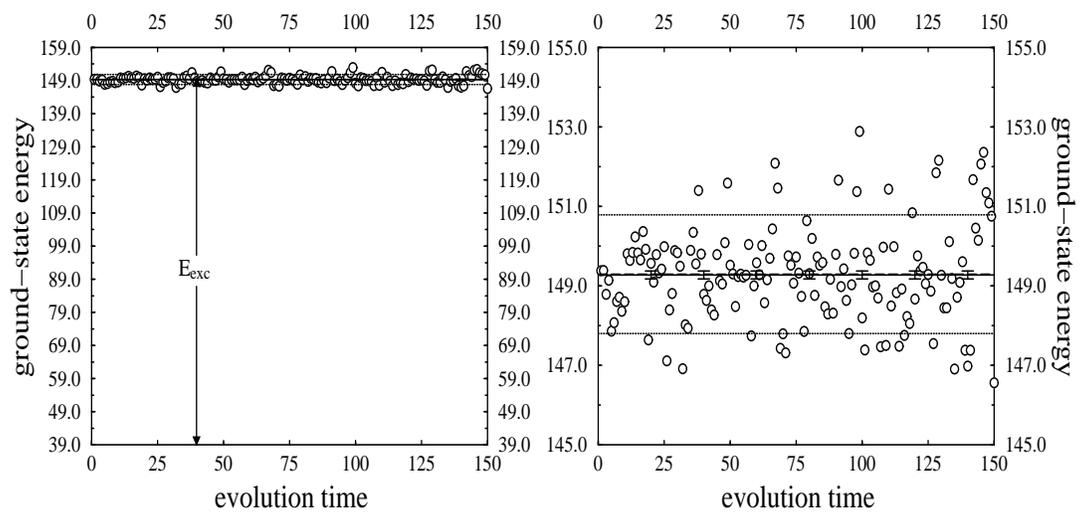
{\bf Fig. 8: }Ground-state energy estimates (circles) for ten
three-dimensional identical fermions (\ref{oscsys}), $\Omega =4$, $\omega =1$
as a function of evolution time on two different scales. The dotted lines at
the right hand side indicate a deviation of one percent from the rigorous
ground-state energy (solid line). The numerically predicted energy is
denoted by the dashed line. The left hand side illustrates the same results
on the scale of exchange contributions.

\end{document}